# Using Graphics Processors for Parallelizing Hash-based Data Carving

Sylvain Collange
*ELIAUS – University of Perpignan*
sylvain.collange@univ-perp.fr

Yoginder S. Dandass
*Mississippi State University*
yogi@cse.msstate.edu

Marc Daumas
*ELIAUS – University of Perpignan*
marc.daumas@univ-perp.fr

David Defour
*ELIAUS – University of Perpignan*
david.defour@univ-perp.fr

**Abstract**

*The ability to detect fragments of deleted image files and to reconstruct these image files from all available fragments on disk is a key activity in the field of digital forensics. Although reconstruction of image files from the file fragments on disk can be accomplished by simply comparing the content of sectors on disk with the content of known files, this brute-force approach can be time consuming.*

*This paper presents results from research into the use of Graphics Processing Units (GPUs) in detecting specific image file byte patterns in disk clusters. Unique identifying pattern for each disk sector is compared against patterns in known images. A pattern match indicates the potential presence of an image and flags the disk sector for further in-depth examination to confirm the match. The GPU-based implementation outperforms the software implementation by a significant margin.*

## 1. Introduction

Discovering known illicit material on digital storage devices is a key component of a digital forensic investigation. Given a large disk that is the subject of a forensic investigation, the basic problem in reconstructing deleted files from fragmented clusters is determining which file clusters contain data from files of interest and the order in which the fragments need to be assembled in order to construct the complete file.

Using existing data carving techniques and tools, it is typically difficult to recover remaining fragments of deleted illicit files whose file system metadata and file headers have been overwritten by newer files. Often, after a file has been deleted, the file system data structures that map the file's content onto clusters on the disk are also reused and cannot be exploited to reconstruct the deleted file. Therefore, tools that reconstruct files (or partial files) by examining individual disk clusters (or sectors) are important to forensic investigators.

A sector-based scan can be used to locate those sectors whose content matches those of sectors from known illicit files. However, brute-force sector-by-sector comparison is prohibitive in terms of time required. Techniques that compute and compare hash-based signatures of sectors in order to filter out those sectors that do not produce the same signatures as sectors from known illicit files are required for accelerating the process.

The approach described in this paper is to extract hashes (*i.e.*, distinct signatures) of sectors from master files (*i.e.*, known files that are the subject of investigation). These signatures are compared with the signatures of the sectors on the disk. When there are millions of master file sectors and millions of disk sectors, a brute-force software-based sequential matching approach takes a large amount of time.

This paper presents an approach that uses Graphics Processing Units (GPUs) to implement parallel pattern matching engines that can significantly speedup the search process. Minimizing the time required to search disk drives is important in situations where investigators have only a short amount of time to scan a disk drive. For example at border crossings and at airports, travelers may only have a short time between connections and law enforcement agencies need to detect evidence of contraband files as quickly as possible. Therefore, it is imperative that the forensic scanning be performed expeditiously. High throughput also becomes essential when the technology described in this article is applied to the monitoring of network traffic, for example in the inspection of traffic going through a digital subscriber line access multiplexer or a cable-modem termination system with a GPU based snort tool [1] that performs hashed based data carving.

The remainder of the paper is organized as follows. Section 2 provides background on hashing, hash-based data carving, GPUs, and related work. Section 3 describes the technical approach, and Section 4 describes the experiments and discusses experimental results. Section 5 concludes with a summary of findings and a discussion of avenues for future research.

## 2. Background

File systems typically store files in clusters of contiguous sectors on disks. For example, in its default configuration, the Windows XP file system, NTFS, allocates space for files in increments of eight contiguous sectors. Each sector is 512 bytes, making each cluster 4096 bytes (*i.e.*, 4KB). There is no guarantee that consecutive clusters belonging to a file will be stored contiguously. When the file system needs to allocate one or more clusters to a file, it allocates clusters from the free space (*i.e.*, the set of clusters that are not currently allocated to other files). Over time, after several cycles of adding and deleting files, the disk's free space becomes fragmented. This fragmentation causes the file system to scatter the content of new files over several non-contiguous clusters.

When a file is deleted, the file system typically marks the file's clusters as being available for reuse but does not *zeroize* (*i.e.*, clear) the content. Therefore, unless a cluster that previously belonged to a deleted file has been overwritten by the contents of another file, it is possible to recover the content of the cluster. Furthermore, in the event that the cluster belonging to the deleted file is allocated to a new file that only overwrites the first few sectors in the cluster, the remaining file fragment data can be examined using the sector signature analysis technique described in this paper. If the file system data structures that map the deleted file's content onto clusters on the disk have not been reused, then the content of deleted files can be recovered relatively easily. Once this file system metadata is overwritten by new files, the recovery of the remaining data fragments becomes more difficult. However, sector signature matching can be applied to any sectors that have not been overwritten by new files.

NTFS stores small files (less than 900 bytes) in the master file table (MFT). The MFT is the file system metadata storage area that contains the file's directory information. Therefore, these small files will not be aligned to disk sector boundaries. However, most files of interest typically range from several kilobytes to gigabytes and are stored on clusters separately from the MFT. Therefore, these large files are guaranteed to be aligned on 512-byte sector boundaries.

### 2.1. Cryptographic and Polynomial Hashing

The MD5 hashing algorithm intended for digital signature applications was described by Rivest in 1992 [2]. MD5 produces a 128-bit (16-byte) hash from data up to ($2^{64}$-1) bits in length. The algorithm pads the input data in order to ensure that the input length is divisible by 512 and proceeds to compute the hash by iteratively processing the input in terms of 512-bit blocks. Although, a number of cryptographic weaknesses have been demonstrated for MD5, it remains an effective hashing algorithm because a small change in input results in significantly large change in the output hash. Therefore, MD5 is a good candidate for developing signatures for sectors.

The SHA-1 hashing algorithm was published as FIPS PUB 180-1 in 1995 and was revised in 2002 as FIPS PUB 180-2 [3]. SHA-1 produces 160-bit (20-byte) hash from data up to ($2^{64}$-1) bits in length. As with the MD5 algorithm, SHA-1 also pads the input data in order to ensure that the input length is divisible by 512 and proceeds to compute the hash by iteratively processing the input in terms of 512-bit blocks. Cryptographic weaknesses have also been reported for SHA-1. However, SHA-1 remains a good candidate for producing sector signatures.

The design of MD5 and SHA-1 algorithms focuses on cryptographic quality, not on facilitating high-speed execution. For example, both algorithms require padding of the input data with a string of zero bits and a 64-bit value containing the original bit-length of the data such that the bit-length of the padded data is divisible by 512 bits. This initial padding is performed regardless of the length of the original data block. However, this initial padding step is not essential for generating signatures of sectors because sectors have a fixed length of 4096 bits. The padding operation will essentially append a block of 512 bits that does not vary based on the content of the block, and therefore, makes no contribution towards producing non-colliding hashes.

Additionally, the MD5 and SHA-1 algorithms specify long sequences of operations that cannot be fully parallelized. This limits the improvements in performance that can be achieved via parallelization of implementations though we could obtain sufficient parallelism on GPUs by operating concurrently on different pieces of data as reported in this work.

SHA-1 is also widely used for password based cryptography. For example PKCS#5 version 2.0 introduced by RSA Laboratories generally uses 4000 iterations of the SHA-1 function to generate an encrypted password from a password and a salt value. This standard is used in Windows password and file encryption, WPA wireless network encryption and many other applications.

CRC32 is a checksum value typically used to validate the correctness of stored data or data transmitted over a network [4]. CRC32 is computed by dividing the input data by the generator polynomial $x^{32} + x^{26} + x^{23} + x^{22} + x^{16} + x^{12} + x^{11} + x^{10} + x^8 + x^7 + x^5 + x^4 + x^2 + x + 1$, using modulo-2 arithmetic. In modulo-2 arithmetic, addition and subtraction resembles the exclusive-or operation (carry processing is not required). The 32-bit (4-byte) remainder of the division operation is the CRC32 value. CRC64 produces a 64-bit (8-byte) checksum value using the generator polynomial $x^{64} + x^4 + x^3 + x + 1$. CRC32 and CRC64 can be used for identification purposes although they are intended for detecting accidental modification of data but not for detecting deliberate modifications.

A number of fast hardware implementations for the MD5 [5], and SHA-1 [6] have been described in the literature. CRC32 and CRC64 implementations can be found on many modern hardware devices that support high-speed serial I/O channels. Hardware implementations have the advantage of computing the hashes at significantly faster rates than possible by software implementations.

### 2.2. Simpler Hashes

The two major problems with cryptographic and CRC hashes are the large number of operations required to compute each iteration of the hashing operation and accessing the hash table (for table-based iterations) when the table is too large to fit in the processor's cache memory. The cryptographic hashes are complex because they need to be irreversible. Fast software implementations of CRC typically require table lookups. Implementing tables requires access to large areas of memory that reduce the performance as this table is randomly accessed. Furthermore, processor caches are of no use and may even slow down the search. A performance breakdown occurs as soon as the lookup tables become too large to fit completely within cache memory.

*djb2* and *sdbm* [7] are two hashing algorithms that operate with no lookup table and do not significantly increase the number of hash collisions in our tests. The *djb2* algorithm, first described by Danier Bernstien in newsgroup *comp.lang.c*, is expressed using the following equation:

$$hash_i = \begin{cases} 5381, & \text{when } i = 0 \\ hash_{i-1} \times 33 + s_i, & \text{when } i > 0 \end{cases} \quad (1)$$

where $hash_i$ is the hash value at the $i^{th}$ iteration and $s_i$ is the input character at iteration $i$ (note 5381 is the hash initialization value before the first iteration is computed). The *sdbm* algorithm is used in the SDBM, and open source database management software, and is expressed using the following equation:

$$hash_i = \begin{cases} 0, & \text{when } i = 0 \\ hash_{i-1} \times 65599 + s_i, & \text{when } i > 0 \end{cases} \quad (2)$$

Implementations of *djb2* and *sdbm* hashing are fast in GPUs because only one multiplication and one addition is required. Furthermore, only the current hash value and the next input character need to be maintained in memory, thereby, minimizing storage requirement. Conversely, these hashes are more sensitive to the number of hash collisions as they increase the ratio of memory access over arithmetic operations.

### 2.3. Data Carving Using Hashes

Dandass *et al.* describe results of a case study in which the SHA1, MD5, CRC64, and CRC32 hashes for over 528 million sectors extracted from over 433 thousand files of different types were analyzed [5]. Particular emphasis was placed on characterizing the viability of using hashes to uniquely identify sectors from JPEG and MD5 files. The results of this study show that MD5 and SHA1 produced no false positive indications. Furthermore, the occurrence of false positives was relatively low for CRC32 and CRC64. Additionally, the smaller size of the CRC32 and CRC64 hashes as compared with SHA1 and MD5 means that a smaller memory capacity is needed in order to deploy a database of hashes of sectors from known contraband files. Reducing memory capacity is an important consideration for the development of a commercially viable forensic took for scanning disks.

Dandass also reports on the results of using disk cluster signatures for detecting contraband data on an FPGA-based platform [8]. However, an FPGA-based implementation provides high-performance at a high cost because FPGAs are expensive. Conversely, GPUs are relatively inexpensive commodity hardware typically found on all workstations and now on high end laptops for agile law enforcement teams

Roussev *et al.* describe techniques for using the djb2 hash function for computing similarity scores for detecting content in files that may be similar to each other [7]. They also present results from case studies parallelizing forensics tools (*e.g.*, Scalpel) for execution on GPUs [9].

### 2.4. Graphics Processors

Graphics hardware has been evolving rapidly over the past several years. Initially, graphics hardware was dedicated to accelerating graphics rendering functionality. Recently, significant developments have made GPUs capable of performing all kinds of general-purpose computations in parallel [10]. This section provides a description of the evolution and capabilities of GPUs.

Original graphics accelerator cards were special-purpose hardware accelerators for graphics programming interfaces such as OpenGL and DirectX. These programming interfaces are used for describing scenes from geometrical objects located by vertices. An image consisting of pixels is rendered using the vertex data. In order to accelerate performance, image rendering functionality is implemented in hardware in the original GPUs as shown in Figure 1.

More recently, operations performed on vertices and pixels have become more flexible through the use of programmable vertex and pixel units called *shaders*. Although vertex and pixel shaders are used for performing different kinds of operations, they share a number of features and offer similar functionality.

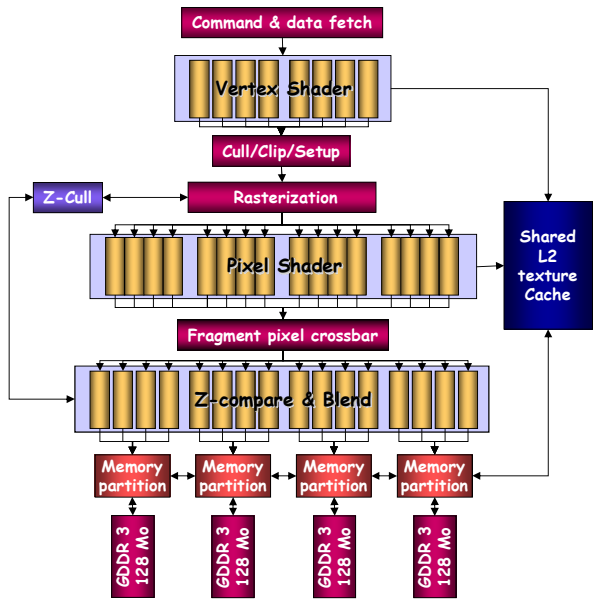

**Figure 1: Organization of early GPUs**

Hardware implementing vertex and pixel shaders include memory units (*e.g.*, texture access units), computational units (*e.g.*, multiply and accumulator units), and specialized hardware (*e.g.*, special function evaluators). In order to efficiently exploit data parallelism, GPUs include a large number of replicated copies of these units. GPUs handle high latency instructions that perform memory accesses by supporting the concurrent execution of thousands of threads.

The DirectX 10 standard and compatible hardware provide a *unified architecture* where vertex and pixel shaders share the same instruction set and processing units. It has been implemented in hardware since the release of the NVidia GeForce 8 and AMD-ATI Radeon HD 2000 generations. An example of such architectures is depicted in Figure 2.

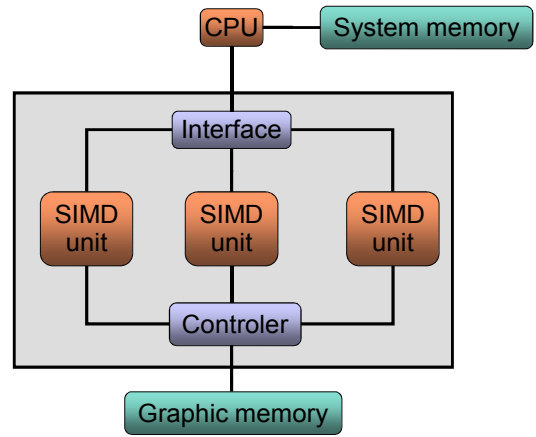

**Figure 2: Organization of NVidia G80**

In this figure it can be seen that the GPU has its own memory which is usually an order of magnitude faster than system memory. The graphic processor is shown as a set of single-instruction multiple-data (SIMD) blocks. Each SIMD block consists of numerous processing elements (PEs). At each clock cycle, all the PEs in a block execute the same instruction in a sequence of instructions, but operate on different data.

These SIMD blocks incorporate a variety of different types of storage such as a set of registers for each PE, memory blocks shared by all the PEs in a SIMD block, constant memory, and read-only texture memory. In addition, the PEs can also read or write global memory available on the graphics card.

SIMD blocks also integrate a variety of computational units in order to implement the functionality offered by the shaders. These include general computation units with embedded multiply-accumulators, texturing and filtering units, and a dedicated unit to evaluate mathematical functions (*e.g.*, sine, cosine, reciprocal, and reciprocal square root), rasterization units, Z-culling units, and interpolation units.

These units can handle integer and floating point arithmetic and there is no overhead associated with mixing both types of operations. Each SIMD block in the GeForce 8 is able to execute a set of 32 floating point additions, multiplications, multiply-and-accumulate or integer additions, bitwise operations, compares, or evaluate the minimum or maximum of 2 numbers in 4 clock cycles. As there is no 32-bit integer multiplication in hardware, evaluating such an operation requires 16 clock cycles for a set.

Figure 3 depicts the NVidia GeForce 8800 GTX. This GPU contains 16 SIMD blocks and each SIMD block is composed of 8 PEs and 2 units to evaluate general functions. It also contains a 768 MB global memory with a peak bandwidth of 103 GB/s which is more than 10 times the available bandwidth between the CPU and the system memory.

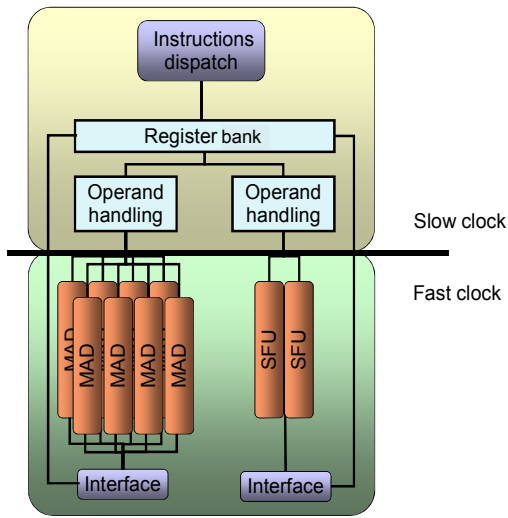

**Figure 3: SIMD unit of NVidia G80**

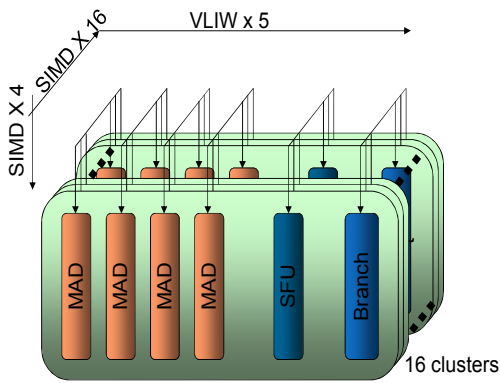

**Figure 4: Organization AMD-ATI R600 cluster**

Figure 4 depicts the AMD-ATI Radeon HD 2900 XT. It contains 4 SIMD blocks, each composed of 16 PEs, capable of performing 5 different instructions simultaneously.

In order to optimize performance, programmers must observe the following two constraints in programming style imposed by the SIMD execution model:
1. Memory contention is minimized by maximizing the use of certain optimized memory access patterns (*i.e.*, coalesced or broadcast accesses described below), depending on the kind of memory being accessed (*i.e.*, shared, constant, or global).
2. Control flow should be the same for all the execution contexts within a SIMD block and divergence of conditional constructs among PEs of the same SIMD block should be minimized.

The first constraint restricts the available memory access patterns based on the type of memory being accessed. There are two primary types of optimized memory access patterns available. *Coalesced* accesses arise from grouping read or write accesses made to consecutive address. For example, if PE 0 requests a read at address $n$, PE 1 requests a read at address $n+1$ and so on, then all reads are performed simultaneously. *Broadcast* accesses arise when all the PEs in an SIMD block read the same address.

The GeForce 8, for example, has a shared constant memory accessible in broadcast mode only, a global memory accessible with coalescing (with additional alignment constraints), and a shared memory that allows coalesced, broadcast, and other patterns. Memory accesses that do not match these patterns are replaced with as many serial accesses as necessary, resulting in decreased performance.

The second constraint means that all PEs in a SIMD block essentially execute the same program synchronously. A jump instruction is executed as a parallel jump only when all the execution contexts follow the same path within a SIMD block, as seen in Figure 5.

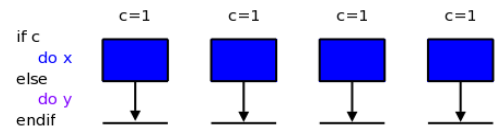

**Figure 5: Execution of a jump instruction when all PEs follow the same path**

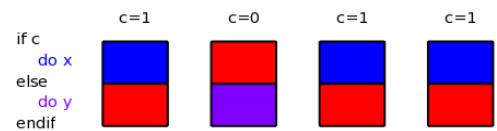

**Figure 6: Execution of a jump instruction when some PEs do not follow the same path**

If any of the branches diverge (*i.e.*, at least one branch within the SIMD block takes a different path) it becomes necessary to execute both branches (in blue and purple in Figure 6) of the conditional construct and a *mask* is applied in order to inhibit writing the results where the condition does not apply (shown in red).

This mechanism of executing both branches in a conditional construct is called *predication*. Some SIMD architectures, including GPUs, are able to dynamically determine at runtime whether a SIMD block can execute only one side of the branch or if it is necessary to execute both sides of the branch and use predication to obtain the final result for all the individual PEs.

## 3. Data Carving Using GPUs

The research presented in this paper explores the feasibility of using the parallelism offered by GPUs to accelerate the detection of sectors from contraband files using sector-level hashes. This application is able to inspect several disk drives simultaneously and asynchronously from each other because computers often contain more than one disk; additionally disks from different computers can be inspected independently.

Because accessing data in memory is time consuming, CPU implementations need to operate in a threaded manner to hide latency. This implementation, instead, uses the ability of GPU to hide memory latency in order to perform an initial filter operation that eliminates sectors from consideration that do not match the signature of sectors from known contraband files. This section describes the architecture of the implementation and describes the various data flow optimizations undertaken in order to maximize performance.

### 3.1. Application Architecture

The overall architecture of the system is described in Figure 7. Sector contents are transferred from the disk drives to the CPU main memory in a two-stage pipelined process using a *double-buffered* implementation for each disk drive. In double buffering, a pair of buffers $\{h_{2k}, h_{2k+1}\}$ in host memory are used in order to optimize throughput of disk $k$. Initially, one disk drive transfers the content of a group of sectors into $h_{2k}$. After this transfer completes, the buffer is marked as ready and the host CPU schedules the simultaneous transfer of the content of another group of sectors from disk $k$ into $h_{2k+1}$.

Meanwhile, ready buffers in CPU memory are transferred to GPU memory following a round robin pattern. As soon as one buffer $h_i$ is completely transferred to GPU it is tagged as free and a transfer is initiated from the appropriate disk to bring more data into main memory.

The transfer of data from CPU to GPU also uses a pair of buffers $\{g_0, g_1\}$ and the processing of the data by the GPU is overlapped with the data transfer operations. While the data is being transferred into $g_0$, the GPU processes data in $g_1$ and similarly when data is being transferred into $g_1$, the GPU processes data in $g_0$. Overlapping the two data transfer operations with the processing at the GPU and performing simultaneously on different hard drives provide significant performance gains over a naïve sequential implementation. Yet, using pairs of buffers imposes a nominal increase in memory requirements.

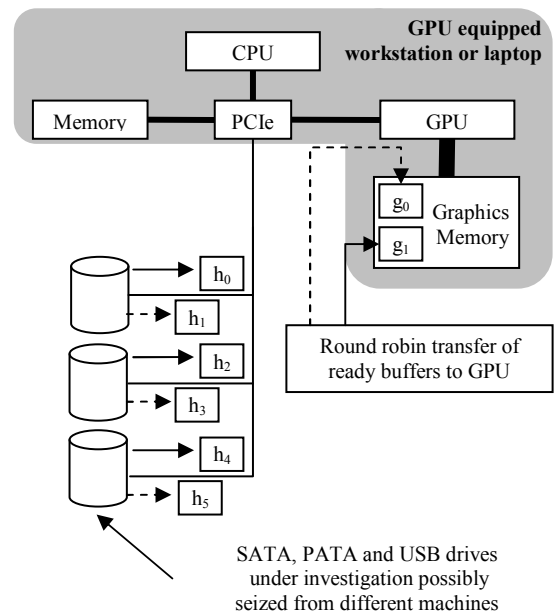

**Figure 7: Architecture of the sector-based signature matching system**

On the GPU, the database of signatures of sectors from known contraband files is organized into hash tables $S_\tau$, $0 \leq \tau \leq 7$. Each hash table, $S_\tau$, contains signatures corresponding to sector $\tau$ in the clusters found on disk drives (note that in this research, clusters are assumed to have the default size of eight sectors). The hash table $S_\tau$ is constructed by successively computing and inserting the hash-based signature of all sectors at position $\tau$ in clusters of known contraband files that are to be detected. The signature is inserted into the hash table at the location, $l$, determined by the $w$ least significant bits in the signature. If there is a hash collision, *linear probing* is used to find an alternative location. Using linear probing, the candidate location for insertion of hash signature *h* into the hash table is given by the following expression:

$$l_c = LSB_w \begin{cases} h, & \text{when } c = 0 \\ l_{c-1} + 1, & \text{when } c > 0 \end{cases} \quad (3)$$

where *c* is the number of collisions that have occurred previously when attempting to insert this hash signature and $LSB_w(x)$ represents the *w* least significant bits of *x*.

At the GPU, signatures of several sectors are computed concurrently and are then looked up in the appropriate hash table. If a matching signature is found, the GPU alerts the host CPU, enabling confirmation (*e.g.*, by performing a byte-for-byte comparison of the suspect sector and the known contraband sector with the matching signature).

### 3.2. Signature Computation on the GPU

Sectors are transferred from the host system's main memory into GPU main memory and stored contiguously in buffers $g_0$ and $g_1$ as shown in figure 8. GPU main memory is external to the GPU chip and needs to be accessed in constrained access patterns in order to improve performance. Straightforward implementations in which threads simultaneously read sector data at offset 0, 1, 2, etc. from main GPU memory results in poor performance because all threads access the same memory bank simultaneously, leading to memory access contention.

| Sector 0 | 0 | 1 | 2 | ... | 14 | 15 | ... |
|---|---|---|---|---|---|---|---|
| Sector 1 | 0 | 1 | 2 | ... | 14 | 15 | ... |
| Sector 2 | 0 | 1 | 2 | ... | 14 | 15 | ... |
| ... | | | | ... | | | |
| Sector 14 | 0 | 1 | 2 | ... | 14 | 15 | ... |
| Sector 15 | 0 | 1 | 2 | ... | 14 | 15 | ... |
| ... | | | | ... | | | |

**Figure 8: Sectors are stored at contiguous addresses in GPU main memory**

In order to accelerate the performance of signature generation and matching, 16x16 blocks of 32-bit words are rearranged in the GPU local memory as depicted in Figure 9 for the first block. This significantly improves the throughput of main GPU memory accesses [11]. Threads are organized in cohorts of 16. First, the 16 threads in a cohort read 16 successive 32-bit words (*i.e.*, a 512 bit span) from one single sector into an on-chip shared memory buffer. The on-chip shared memory is organized in a columnar manner in hardware such that successive words appear in different banks. The 16 threads can simultaneously access the memory provided that the threads access different banks as shown in Figure 9. Note that the read operation is a coalesced memory access to main GPU memory and the write operation takes place in parallel for all the 16 threads. These operations are repeated 16 times until 512-bit portions of 16 sectors have been uploaded to the shared on-chip memory buffer.

| Sector 0 | 0 | 1 | 2 | ... | 14 | 15 |
|---|---|---|---|---|---|---|
| Sector 1 | 15 | 0 | 1 | ... | 13 | 14 |
| Sector 2 | 14 | 15 | 0 | ... | 12 | 13 |
| ... | | | | ... | | |
| Sector 14 | 2 | 3 | 4 | ... | 0 | 1 |
| Sector 15 | 1 | 2 | 3 | ... | 15 | 0 |

**Figure 9: Data are first rearranged in GPU shared memory by group of 16x16 words**

The cohort of 16 threads computes the partial signature from the first 512 bits of data from 16 sectors (*i.e.*, thread *q* in the cohort computes the signature for sector *q*). This process is repeated eight times as the threads fetch more data from the 16 sectors and continue the computation of partial signatures until complete signatures of the 16 sectors have been computed.

Each of the 16 threads next performs a hash table lookup of the signatures in the set of hash tables in main memory. Linear probing is used in order to resolve collisions. Essentially, if the entry at the location indicated by the computed hash value is occupied by a non-null signature that does not match the computed signature, the thread looks at the next location. This process continues until the signature is found in the hash table or a null entry is found. When the signature is found in the hash table, this indicates a positive match. Conversely, if a null hash table entry is found first, this indicates that the signature does not exist in the table of signatures of known contraband sectors.

After writing information about positive matches to a memory buffer where the host CPU can access the match indications, the cohort of threads begins transferring and processing the next set of 16 sectors from main GPU memory.

In order to fully exploit the parallelism available in modern GPUs, the implementation allows several such cohorts to be executed in parallel. The exact number of cohorts that can be executed simultaneously depends on the platform used for executing the system. For the hardware used in the experiments reported in this paper, the best performance is achieved with eight cohorts. However, this number is expected to vary with future generations of GPUs.

## 4. Experimental Results

In order to determine the effectiveness of using GPUs for computing and matching hashes, two different sets of experiments were conducted. The first set of experiments replicates the hash quality experiments conducted by Dandass *et al.* in order to determine the number of collisions resulting from using the signatures based on the *djb2* hashing function. Note that a 64-bit accumulator was used in the *djb2* hash computation, resulting in 64-bit wide hashes.

The second set of experiments measures the execution performance of the sector signature scanning implementation on the Nvidia GeForce 8500 GT GPU and Intel Xeon 5140 2.33 GHz CPU using CRC64- and *djb2*-based signatures. The CPU-based implementation is sequential while the GPU-based implementation employs 2048 threads (*i.e.*, 128 cohorts).

A set of 105,506 unique files consisting of 86,655,112 unique sectors of random data were used in this experiment. Eight hash tables occupying 32 MB and indexed by the 22 least significant bits of the signature were used for storing the database of known contraband hashes. The second set of experiments has also been ported to a laptop equipped with a NVidia GeForce 8600M GT. The performance of the 8600M is slightly higher than the performance of the NVidia GeForce 8400 GT on the workstation.

### 4.1. Hash Collisions Using *djb2*

Only 32 of the *djb2* hashes from these sectors collided with the *djb2* hashes from other sectors (*i.e.*, the collision rate was 0.33%). This indicates that djb2 is a relatively good candidate for generating hash-based signatures because the potential number of false-positive indications is small. Although fewer files were used in this experiment that those used by Dandass et *al.*, based on the evidence of other research and the 64-bit width of the hash, it is anticipated that *djb2* will perform well in general.

### 4.2. System Performance on the GPU

Figures 10 and 11 compare the performance of various GPU-based implementations and a CPU-based implementation using CRC64 function for signature generation. The X-axis represents the occupancy of the hash table and the Y-axis represents the throughput of the implementation in megabytes per second. The CPU-based implementation has the lowest performance as expected because of its sequential nature.

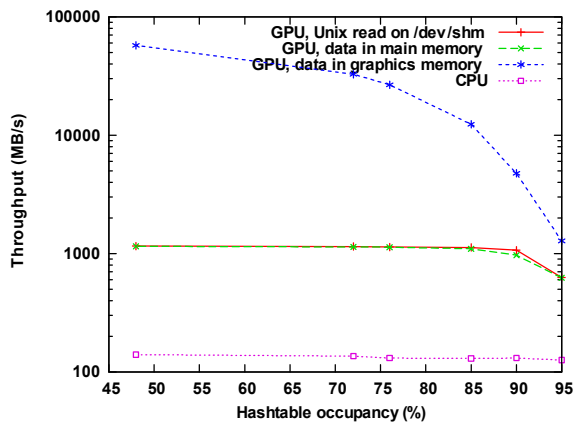

**Figure 10: CRC64-based implementation on NVidia GeForce 9800 GX2**

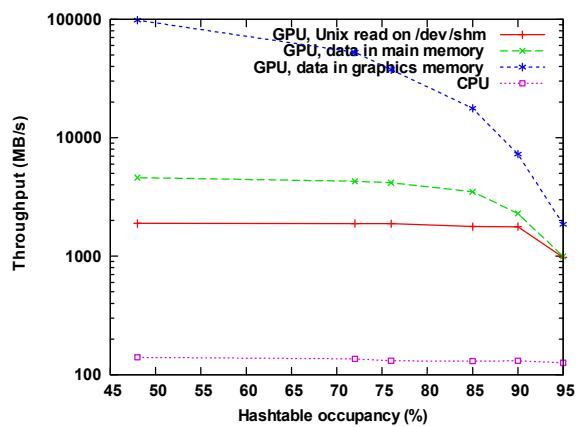

**Figure 11: CRC64-based implementation on an NVidia GT 200 engineering sample**

Three GPU-based implementations are plotted in increasing order of performance:
1. Data in a virtual disk drive in memory /dev/shm – for this implementation, the files to be examined are placed on a virtual disk drive memory in order to eliminate the effects of the relatively slow disk I/O.
2. Data in main memory – in this implementation, the data is placed in the workstation's main

memory in order to eliminate the overhead from the file I/O subsystem.
3. Data in graphics memory – in this implementation, the sector data is assumed to be in main GPU memory in order to eliminate the overhead of all data transfer operations (*i.e.*, this is a measure of raw GPU performance).

These plots clearly show that the large-scale parallelism available on the GPU enables the fast execution of signature generation and matching. However, I/O is a significant bottleneck.

Performance decreases with increasing hash table occupancy because of the increasing number of linear probing operations required. However, even with 90% hash table occupancy, throughput is over 1700 MB/s when file system and disk I/O overheads are not considered. When file system I/O is included, the GPU's performance remains at 675 MB/s at up to 90% hash table occupancy.

Figure 12 plots the performance of the *djb2*-based implementation. This implementation has slightly better performance because the *djb2* hash function is simpler to compute than CRC64 on GPUs. Note that for the higher hash-table occupancies, the linear probe functionality imposes a significant overhead. In his FPGA-based implementation, Dandass used a binary search mechanism; this may be a feasible alternative to use in implementations with high hash-table occupancy.

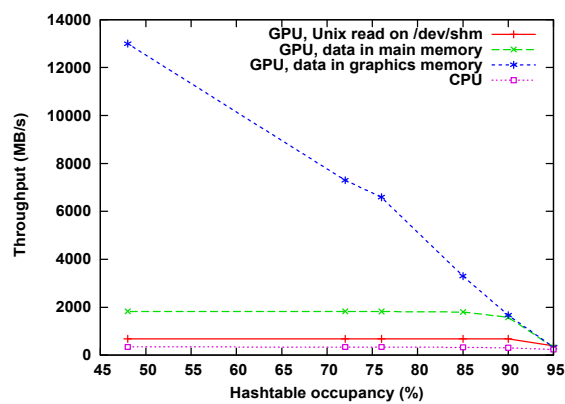

**Figure 12: *djb2*-based implementation on NVidia GeForce 8500 GT**

For CRC64 and djb2, the GPU implementation always performs better than the CPU implementation. Several factors contribute to this performance differential. First, GPUs have much higher computational power than CPU. Although the CPU implementation used in these experiments is not optimized (because one iteration of CRC64 takes 17 cycles per byte whereas the implementation by Sarwate takes 6.67 cycles per byte [12]), we have a 13-fold improvement in performance between the GPU and CPU implementations. Additionally, with big hash tables that do not fit entirely in CPU cache, the data cache of the CPU adds extra cycles to access data located in main memory on a bus shared with all active processes. Conversely, in the GPU implementation, the hash table is entirely located in high bandwidth graphics memory dedicated to the GPU that is directly accessed without any intervening cache. In addition, the threaded GPU implementation is naturally better at hiding access data latency.

These results show that it is feasible to design an implementation with the computational capacity to simultaneously scan up about six separate disk drives each transferring data a peak rate of over 100 MB/s. In the future, it may be possible to improve peak performance by eliminating certain OS overheads incurred for transferring data between the disk and the host. However, in the implementation utilizing a single disk drive, throughput is limited to 55MB/s (this is essentially, the sustained data rate of the disk drive used in the experimental platform).

## 5. Conclusions and Future Work

This paper presents a GPU-based approach for rapidly scanning the content of sectors of disks in order to look for fragments of possibly deleted image files. Hash-based signatures of disk sectors are compared with signatures of sectors from known contraband files and matches are flagged for further investigation. Although the fundamental file fragment signature scanning techniques presented here can be implemented using conventional software techniques, the rapidly increasing storage capacity of disk drives is necessitating the development of faster alternatives.

Experimental results reported in this paper demonstrate the viability of using GPU technology for scanning hard disks for fragments of known image files. GPUs are a commodity item and are found in most modern workstations and in high end laptops. This means that the relatively inexpensive parallel processing capabilities of the GPU can be readily utilized in an ordinary workstation platform or in nomad applications, without requiring costly FPGA hardware used in prior research.

Near term future research will focus on improving the performance of the signature matching codes on the GPU and on overlapping disk I/O with GPU I/O and GPU processing in order to minimize latency. Clearly, reading an entire disk drive at peak data transfer capacity can take a long time (*e.g.*, the Seagate

Barracuda 7200.11 500GB disk drive will take approximately one hour and twenty minutes to read using the peak sustained data rate of 105 MB/s). Therefore, future research will also focus on developing techniques for determining those areas on disk that are likely to contain prohibited data (*e.g.*, using file system metadata). This will enable a focused search of disk drives given small windows of investigation time.

Longer term research will focus on developing a workstation- and a laptop-based platform incorporating a high-end GPU that will scan for signatures of known contraband files for several disk drives simultaneously and will allow the operator to dynamically attach disk drives using USB, PATA, and SATA interfaces. This capability will enable the inspection without requiring the system to be power cycled in order to remove or attach new drives. The authors are working with a startup company in France with the goal of developing such a platform.

The computing power exhibited by GPUs in this work opens new perspectives on inspections that were long thought to be unreachable. For example, the GPU implementations described in this paper are able to process every 64-byte fragment aligned on 32-bit boundaries in disks at a sustained rate of about 500 MB/s. Law enforcement agencies can therefore search for small images and logos embedded in documents and files in archives provided they are aligned to 32-bit boundaries.

## 6. Acknowledgements

This work is supported in part by NSF Cyber Trust grant SCI-0430354 and in part by donations of software and hardware by NVidia and AMD. The work was partially performed while Yoginder S. Dandass visited the University of Perpignan Via Domitia as an invited professor, supported, in part, by the French Région Languedoc-Roussillon. The authors also express their gratitude to inspectors from the Direction de la Surveillance du Territoire, the French counter-intelligence and homeland security agency.

## 7. References

1. N-F. Huang, H-W. Hung, S-H. Lai, Y-M. Chu, and W-Y. Tsai, "A GPU-Based Multiple-Pattern Matching Algorithm for Network Intrusion Detection Systems" in *22nd International Conference on Advanced Information Networking and Applications*, 2008.

2. R. L. Rivest, "The MD5 Message Digest Algorithm," Internet RFC 131, 1992, retrieved January 18, 2008, from http://people.csail.mit.edu/rivest/Rivest-MD5.txt

3. NIST-ITL, "Secure Hash Standard (SHS)," 2002, Retrieved January 18, 2008 from http://csrc.nist.gov/publications/fips/fips180-2/fips180-2.pdf.

4. D. V. Sarwate, "Computation of Cyclic Redundancy Checks via Table Look-Up," *Communications of the ACM*, 31(8), 1988.

5. Y. S. Dandass, N. J. Necaise, and S. R. Thomas, "An Empirical Analysis of Disk Sector Hashes for Data Carving," *Journal of Digital Forensic Practice*, 2(2), 2008

6. J. Deepakumara, H. M. Heys, and R. Venkatesan, "FPGA implementation of MD5 hash algorithm," in *Proceedings of the Canadian Conference on Electrical and Computer Engineering*, 2001.

7. V. Roussev, G. G. Richard III, and L. Marziale, "Multi-resolution similarity hashing," *Digital Investigation*, 4(S1), 2007.

8. Y. S. Dandass, "Hardware-assisted Scanning for Signature Patterns in Image File Fragments," in *Proceedings of the 40th Hawaii International Conference on System Sciences*, 2007.

9. L. Marzialea, G. G. Richard III, and V. Roussev, "Massive threading: Using GPUs to increase the performance of digital forensics tools," 4(S1), 2007.

10. S. Collange, M. Daumas and D. Defour, "Line-by-line spectroscopic simulations on graphics processing units", *Computer Physics Communications*, 178(2), 2008.

11. E. Lindholm, J. Nickolls, S. Oberman, and J. Montrym, "NVidia Tesla: A unified graphics and computing architecture", *IEEE Micro,* 28(2).

12. M. E. Kounavis and F. L. Berry, "Novel Table Lookup-Based Algorithms for High Performance CRC Generation", *IEEE Transaction on Computers*, 20 May, 2008.

13. A. P. Kakarountas, H. Michail, A. Milidonis, C. E. Goutis, and G. Theodoridis., "High-Speed FPGA Implementation of Secure Hash Algorithm for IPSec and VPN Applications," *Journal of Supercomputing*, 37(2).